\documentstyle[multicol,aps,psfig]{revtex}
\renewcommand{\narrowtext}{\begin{multicols}{2}
\global\columnwidth20.5pc\noindent}
\renewcommand{\widetext}{\end{multicols}
\global\columnwidth42.5pc}
\begin{document}
\draft
\preprint{1 March 2004}
\title{Soliton-induced optical absorption of halogen-bridged mixed-valence
       \\
       binuclear metal complexes}
\author{Jun Ohara and Shoji Yamamoto}
\address{Division of Physics, Hokkaido University,
         Sapporo 060-0810, Japan}
\date{Received 1 March 2004}
\maketitle
\begin{abstract}
Employing the one-dimensional single-band extended Peierls-Hubbard model,
we investigate optical conductivity for solitonic excitations in
halogen-bridged binuclear metal ($M\!M\!X$) complexes.
Photoinduced soliton absorption spectra for $M\!M\!X$ chains possibly
split into two bands, forming a striking contrast to those for
conventional mononuclear metal ($M\!X$) analogs, due to the broken
electron-hole symmetry combined with relevant Coulomb and/or
electron-phonon interactions.
\end{abstract}
\narrowtext

\section{Introduction}

   Halogen ($X$)-bridged mixed-valence metal ($M$) complexes
\cite{G6408,O2023} such as
[Pt(en)$_2X$](ClO$_4$)$_2$
($X=\mbox{Cl},\mbox{Br},\mbox{I}$;
 $\mbox{en}=\mbox{ethylenediamine}
 =\mbox{C}_2\mbox{H}_8\mbox{N}_2$),
which are referred to as $M\!X$ chains, have been playing a prominent role
in understanding the electronic properties of one-dimensional
Peierls-Hubbard systems.
The competing electron-electron and electron-phonon interactions yield
various ground states \cite{N3865,B13228}, which can be tuned by chemical
substitution \cite{O9} and pressure \cite{G1191}.
Solitonic excitations inherent in charge-density-wave (CDW) ground states
stimulate further interest in $M\!X$ complexes.
In this context, we may be reminded of polyacetylene, the trans isomer of
which exhibits topological solitons \cite{S1698,T2388}.
Several authors \cite{I137,O250,B339} had an idea of similar defect states
existing in $M\!X$ chains.
Photogenerated solitons \cite{K2122,O2248} were indeed observed in
the Pt$X$ compounds.

   In recent years, binuclear metal analogs of $M\!X$ complexes, which
are referred to as $M\!M\!X$ chains, have attracted further interest
exhibiting a wider variety of ground states
\cite{K435,Y125124}, successive thermal phase transitions
\cite{K10068,Y1198}, photo- and pressure-induced phase transitions
\cite{S1405,Y140102,M046401}, and incomparably larger room-temperature
conductivity \cite{K1931}.
In such circumstances, soliton solutions of an $M\!M\!X$ Hamiltonian of
the Su-Schrieffer-Heeger type \cite{S1698} have recently been investigated
both analytically and numerically \cite{Y189}.
The direct $M\!-\!M$ overlap contributes to the reduction of the effective
on-site Coulomb repulsion and therefore electrons can be more itinerant in
$M\!M\!X$ chains.
Hence we take more and more interest in solitons as charge or spin
carriers.

   The ground-state properties of $M\!M\!X$ complexes were well revealed
by means of X-ray diffraction \cite{J1415}, nuclear magnetic resonance
\cite{K40}, and the Raman and M\"ossbauer spectroscopy \cite{K10068},
while very little \cite{W1195,K2163} is known about their excitation
mechanism.
{\it Photoinduced} absorption spectra, which served as prominent probes
for nonlinear excitations in $M\!X$ complexes \cite{K2122,O2248}, have,
to our knowledge, not yet measured on $M\!M\!X$ complexes probably due to
the lack of a guiding theory.
Thus motivated, we study optical conductivity for $M\!M\!X$ solitons with
particular emphasis on a contrast between topical $M\!M\!X$ and
conventional $M\!X$ complexes.
Photoexcited $M\!M\!X$ chains may yield distinct spectral shapes due to
the definite breakdown of the electron-hole symmetry.

\section{Model Hamiltonians and Their Ground-State Properties}

   We describe $M\!X$ and $M\!M\!X$ chains by the one-dimensional
$\frac{1}{2}$- and $\frac{3}{4}$-filled single-band Peierls-Hubbard
Hamiltonians
\begin{mathletters}
   \begin{eqnarray}
      &&
      {\cal H}_{M\!X}=
       h_{M\!X\!M}^{(1,1)}+h_{M}^{(1,1)}+h_{M\!X},
      \label{E:HMX}\\
      &&
      {\cal H}_{M\!M\!X}=
       h_{M\!M}^{(1,2)}+h_{M\!X\!M}^{(1,2)}+h_{M}^{(1,2)}+h_{M\!X},
      \label{E:HMMX}
   \end{eqnarray}
   \label{E:H}
\end{mathletters}
$\!\!\!\!$
respectively, where
\begin{eqnarray}
   &&
   h_{M\!M}^{(\mu,\nu)}
  =-t_{M\!M}\sum_{n,s}
    (a_{\mu:n,s}^\dagger a_{\nu:n,s}+a_{\nu:n,s}^\dagger a_{\mu:n,s})
   \nonumber\\
   &&\qquad\quad
   +V_{M\!M}\sum_{n,s,s'}n_{\mu:n,s}n_{\nu:n,s'},
   \\
   &&
   h_{M\!X\!M}^{(\mu,\nu)}
  =-\sum_{n,s}
    \bigl[t_{M\!X\!M}-\alpha(l_{n+1}^{(-)}+l_{n}^{(+)})\bigr]
    (a_{\mu:n+1,s}^\dagger a_{\nu:n,s}
   \nonumber\\
   &&\qquad\quad
    +a_{\nu:n,s}^\dagger a_{\mu:n+1,s})
   +V_{M\!X\!M}\sum_{n,s,s'}n_{\mu:n+1,s}n_{\nu:n,s'},
   \\
   &&
   h_{M}^{(\mu,\nu)}
  =-\beta\sum_{n,s}
    (l_{n}^{(-)}n_{\mu:n,s}+l_{n}^{(+)}n_{\nu:n,s})
   \nonumber\\
   &&\qquad\quad
   +\frac{\nu U_{M}}{2}\sum_{n}
    (n_{\mu:n,+}n_{\mu:n,-}+n_{\nu:n,+}n_{\nu:n,-}),
   \\
   &&
   h_{M\!X}
  =\frac{K_{M\!X}}{2}\sum_{n}
   \bigl[(l_{n}^{(-)})^2+(l_{n}^{(+)})^2\bigr].
\end{eqnarray}
Here, $n_{\mu:n,s}=a_{\mu:n,s}^\dagger a_{\mu:n,s}$ with
$a_{\mu:n,s}^\dagger$ being the creation operator of an electron with spin
$s=\pm\frac{1}{2}$ (up and down) for the $M$ $d_{z^2}$ orbital labeled as
$\mu=1,2$ in the $n$th $M\!X$ or $M\!M\!X$ unit,$\,$
$t_{M\!M}$$\,$ and$\,$ $t_{M\!X\!M}$$\,$ describe$\,$ the$\,$ intra-$\,$
and
\begin{figure}
\centerline
{\mbox{\psfig{figure=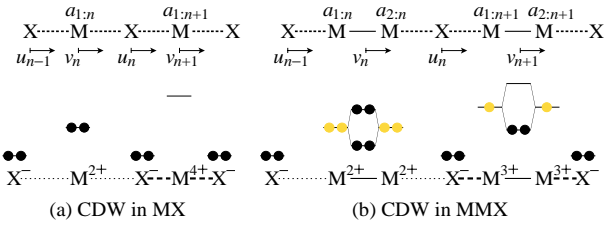,width=80mm,angle=0}}}
\vspace*{1mm}
\caption{Schematic representation of CDW ground states
         observed in $M\!X$ (a) and $M\!M\!X$ (b) complexes.}
\label{F:CDW}
\end{figure}
\noindent
interunit
electron hoppings, respectively,
$\alpha$ and $\beta$ are the site-off-diagonal and site-diagonal
electron-lattice coupling constants, respectively,
$l_{n}^{(-)}=v_n-u_{n-1}$ and $l_{n}^{(+)}=u_n-v_n$ with $u_n$ and $v_n$
being, respectively, the chain-direction displacements of the halogen and
metal in the $n$th unit from their equilibrium position, and $K_{M\!X}$ is
the metal-halogen spring constant.
We assume, based on the thus-far reported experimental observations,
that every $M_2$ moiety is not deformed.
The notation is further explained in Fig. \ref{F:CDW}.
We set $t_{M\!X\!M}$ and $K_{M\!X}$ both equal to unity in the following.

   The existent $M\!M\!X$ complexes consist of two families:
 $R_4$[Pt$_2$(pop)$_4X$]$\cdot$$n$H$_2$O
($X=\mbox{Cl},\mbox{Br},\mbox{I}$;
 $R=\mbox{Li},\mbox{K},\mbox{Cs},\cdots$;
 $\mbox{pop}=\mbox{diphosphonate}
 =\mbox{P}_2\mbox{O}_5\mbox{H}_2^{\,2-}$)
 and
 $M_2$(dta)$_4$I
($M=\mbox{Pt},\mbox{Ni}$;
 $\mbox{dta}=\mbox{dithioacetate}=\mbox{CH}_3\mbox{CS}_2^{\,-}$),
which are referred to as pop and dta complexes.
The pop complexes possess the mixed-valence ground
state illustrated in Fig. \ref{F:CDW}(b) \cite{K40,B1155}, while the dta
complexes exhibit distinct types of ground states
\cite{K435,Y125124,K10068,B4562} due to the predominant site-off-diagonal
electron-phonon coupling \cite{B444} or metal-on-site Coulomb repulsion
\cite{B2815}.
Conventional $M\!X$ complexes structurally resemble $M\!M\!X$ complexes of
the pop type and have the mixed-valence ground state illustrated in
Fig. \ref{F:CDW}(a) \cite{O2023}, which is symmetrically equivalent to
that in Fig. \ref{F:CDW}(b) \cite{Y125124,Y422}.
Hence we study the pop complexes in our first attempt to compare the
photoproducts in $M\!M\!X$ and $M\!X$ chains.
While the dta complexes behave as $d$-$p$-hybridized two-band materials in
general, the pop complexes are well describable within
$d_{z^2}$-single-band Hamiltonians \cite{Y125124}.
Since the site-off-diagonal electron-phonon coupling is of little
significance with the CDW backgrounds of the Fig. \ref{F:CDW} type, we set
$\alpha$ equal to zero.
The rest of parameters are, unless otherwise noted, taken as
$U_M=1.2$, $V_{M\!X\!M}=0.3$, and $\beta=0.7$ for $M\!X$ chains
\cite{G6408,I1380,T2212}, while
$U_M=1.0$, $V_{M\!M}=0.5$, $V_{M\!X\!M}=0.3$, and $\beta=1.4$ for
$M\!M\!X$ chains \cite{Y140102,K2163}.

\section{Calculational Procedure}

   We treat the Hamiltonians (\ref{E:H}) within the Hartree-Fock
approximation.
The lattice distortion is adiabatically determined and can therefore be
expressed in terms of the electronic density matrices
$\langle n_{\mu:n,s}\rangle_{\rm HF}$,
where $\langle\cdots\rangle_{\rm HF}$ denotes the thermal average over
the Hartree-Fock eigenstates.
Since no spontaneous deformation of the metal sublattice has been
observed in any $M\!X$ and $M\!M\!X$ pop complexes, we enforce the
constraint $l_{n+1}^{(-)}+l_{n}^{(+)}=0$ on every $M\!-\!X\!-\!M$ bond.
Then the force equilibrium conditions
\begin{equation}
   \frac{\partial\langle{\cal H}_{M\!X} \rangle_{\rm HF}}
        {\partial l_n^{(\pm)}}=0,\ \ 
   \frac{\partial\langle{\cal H}_{M\!M\!X}\rangle_{\rm HF}}
        {\partial l_n^{(\pm)}}=0,
\end{equation}
yield
\begin{equation}
   \begin{array}{lll}
     2K_{M\!X}l_n^{(-)}
     &=&\beta{\displaystyle\sum_s}
        (\langle n_{\mu:n,s}\rangle_{\rm HF}
        -\langle n_{\nu:n-1,s}\rangle_{\rm HF}),\\
     2K_{M\!X}l_n^{(+)}
     &=&\beta{\displaystyle\sum_s}
        (\langle n_{\nu:n,s}\rangle_{\rm HF}
        -\langle n_{\mu:n+1,s}\rangle_{\rm HF}),\\
   \end{array}
\end{equation}
where $\mu=\nu=1$ for $M\!X$ chains, while $\mu=1$, $\nu=2$ for $M\!M\!X$
chains.

   The real part of the optical conductivity is given by
\begin{equation}
   \sigma(\omega)
    =\frac{\pi}{N\omega}\sum_{\epsilon,\epsilon'}
     f(\epsilon)\bigl[1-f(\epsilon')\bigr]
   \nonumber\\
   |\langle\epsilon'|{\cal J}|\epsilon\rangle|^2
   \delta(\epsilon'-\epsilon-\hbar\omega),
\end{equation}
where $f(\epsilon)=(e^{\epsilon/k_{\rm B}T}+1)^{-1}$ and
$\langle\epsilon'|{\cal J}|\epsilon\rangle$ is the matrix element of the
current density operator ${\cal J}$ between the eigenstates of energy
$\epsilon$ and $\epsilon'$.
${\cal J}$ is defined as
\begin{equation}
   {\cal J}_{M\!X}=j_{M\!X\!M}^{(1,1)},\ \ 
   {\cal J}_{M\!M\!X}=j_{M\!M}^{(1,2)}+j_{M\!X\!M}^{(1,2)},
\end{equation}
for $M\!X$ and $M\!M\!X$ chains, respectively, with
\begin{eqnarray}
   &&
   j_{M\!M}^{(\mu,\nu)}
    =\frac{ie}{\hbar}c_{M\!M}t_{M\!M}\sum_{n,s}
     (a_{\nu:n,s}^\dagger a_{\mu:n,s}-a_{\mu:n,s}^\dagger a_{\nu:n,s}),
   \\
   &&
   j_{M\!X\!M}^{(\mu,\nu)}
    =\frac{ie}{\hbar}c_{M\!X\!M}\sum_{n,s}
     \bigl[t_{M\!X\!M}-\alpha(l_{n+1}^{(-)}+l_{n}^{(+)})\bigr]
   \nonumber\\
   &&\qquad\quad\times
     (a_{\mu:n+1,s}^\dagger a_{\nu:n,s}-a_{\nu:n,s}^\dagger a_{\mu:n+1,s}),
\end{eqnarray}
where $c_{M\!M}$ and $c_{M\!X\!M}$ are the average $M\!-\!M$ and
$M\!-\!X\!-\!M$ distances, respectively, and are set for
$c_{M\!X\!M}=2c_{M\!M}$ \cite{C4604,C409}.

   In order to elucidate the intrinsic excitation mechanism, solitons are
calculated at a sufficiently low temperature without any assumption on
their spatial configurations.
Charged solitons (S$^{\sigma}$; $\sigma=\pm$) are obtained by setting the
numbers of up- and down-spin electrons, $N_+$ and $N_-$, both equal to
$(p-1/2)N-\sigma/2$, while spin-$s$ neutral solitons
[S$^{0s}$; $s=\pm\frac{1}{2}$] with $N_\pm=(p-1/2)N\pm s$, where
$N$, taken to be $401$ in our calculation, is the number of unit cells and
$p$ is the number of metal sites in the unit cell.
\begin{figure}
\centerline
{\mbox{\psfig{figure=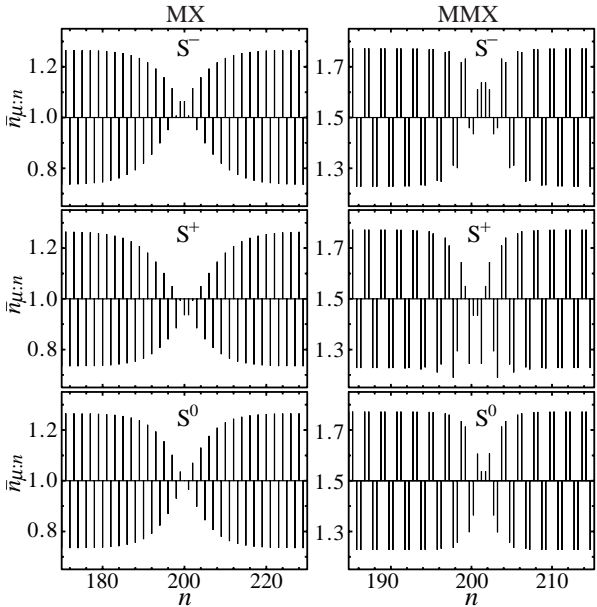,width=80mm,angle=0}}}
\caption{Spatial configurations of the optimum solitons in $M\!X$ and
         $M\!M\!X$ chains.}
\label{F:config}
\end{figure}
\begin{figure}
\centerline
{\mbox{\psfig{figure=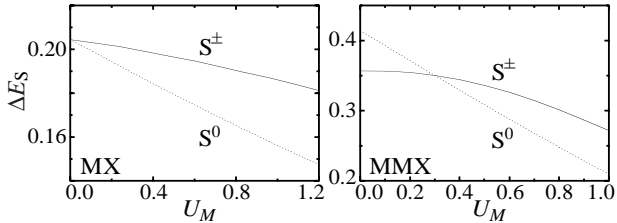,width=82mm,angle=0}}}
\caption{The formation energies of the stably located solitons as
         functions of the Coulomb interaction, where
         $\beta=0.7$ and $U_M$ varies keeping the relation
         $U_M=V_{M\!X\!M}/0.25$ for $M\!X$ chains, while
         $\beta=1.4$ and $U_M$ varies keeping the relation
         $U_M=V_{M\!M}/0.5=V_{M\!X\!M}/0.3$ for $M\!M\!X$ chains.}
\label{F:Ef}
\end{figure}

\section{Results}

   We show the spatial configurations of the optimum solitons in
Fig. \ref{F:config} and plot their formation energies $\Delta E_{\rm S}$
in Fig. \ref{F:Ef}.
In $M\!X$ chains, charged solitons are stable laying their centers on
halogen sites, while neutral solitons on metal sites.
In $M\!M\!X$ chains, all the optimum solitons are halogen-centered.
As the Peierls gap $E_{\rm gap}$ increases, solitons generally possess
increasing energies and decreasing extents, and end up with immobile
defects.
$U_M$ and $V_{M\!M}$ reduce $E_{\rm gap}$ and thus enhance the mobility
of solitons, while $V_{M\!X\!M}$ has an opposite effect.
Without any Coulomb interaction, the soliton formation energies are all
degenerate and scaled by the Peierls gap as
$\Delta E_{\rm S}=E_{\rm gap}/\pi$ in the weak-coupling region
$E_{\rm gap}\alt t_{M\!X\!M}$ \cite{Y165113}, whereas their degeneracy is
lifted and neutral solitons have higher energies than charged solitons
with further increasing gap.
Since $E_{\rm gap}$ decreases with increasing $U_M$, $\Delta E_{\rm S}$ is
a decreasing function of $U_M$.
Neutral solitons are more sensitive to the Coulomb interaction and thus
their formation energy turns smaller than that of charged solitons with
increasing $U_M$.
Considering that $E_{\rm gap}$ is a linear function of $U_M$, we learn
that neutral solitons well keep the scaling relation
$\Delta E_{\rm S}\propto E_{\rm gap}$ against the Coulomb interaction.

\begin{figure}
\centerline
{\mbox{\psfig{figure=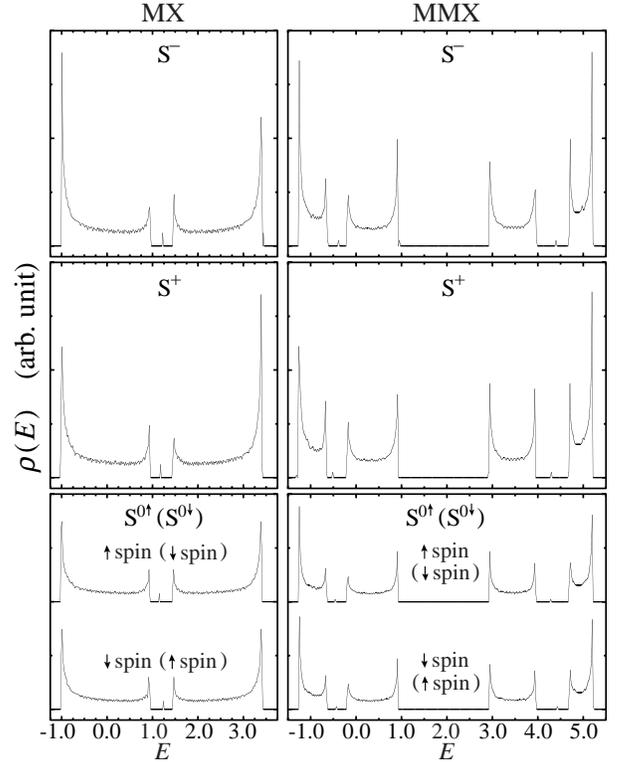,width=80mm,angle=0}}}
\caption{Density of states for the optimum soliton solutions.}
\label{F:DOS}
\end{figure}

   Figure \ref{F:config} suggests that an $M\!X$ soliton of charge
$\sigma$ and spin $s$, S$^{\sigma s}$, described in terms of electrons is
equivalent to its counterpart S$^{-\sigma -s}$ described in terms of
holes.
Such a symmetry is more directly observed through the energy
structures shown in Fig. \ref{F:DOS}.
Solitons generally exhibit an additional level within the gap.
There appear further soliton-related$\,$ levels$\,$ in$\,$ the$\,$
strong-coupling$\,$ region$\,$ \cite{T4074}.
\begin{figure}
\centerline
{\mbox{\psfig{figure=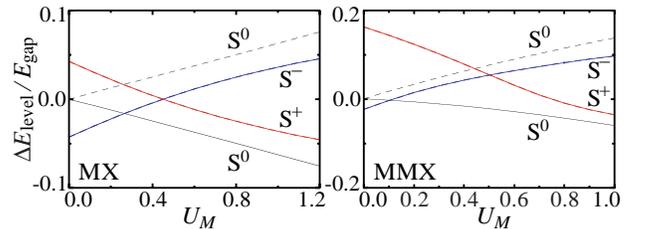,width=82mm,angle=0}}}
\caption{(Color online)
         Energy shifts of the localized soliton levels from the
         gap center scaled by the Peierls gap as functions of the
         Coulomb interaction.
         The parametrization is the same as that in Fig. \ref{F:Ef}.
         The level structures of up- and down-spin electrons are plotted
         by solid (broken) and broken (solid) lines, respectively, for
         S$^{0\uparrow}$ (S$^{0\downarrow}$), while they are degenerate
         for charged solitons.}
\label{F:level}
\end{figure}

\begin{figure}
\centerline
{\mbox{\psfig{figure=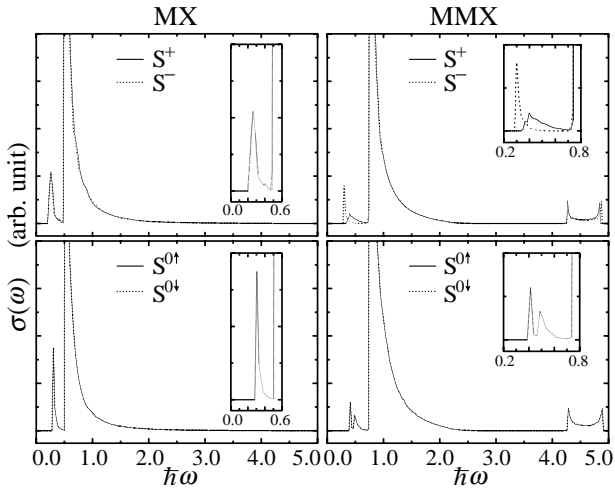,width=82mm,angle=0}}}
\caption{Absorption spectra with a soliton excited for $M\!X$ and
         $M\!M\!X$ chains.
         The low-frequency structures are scaled up in insets.}
\label{F:OC}
\end{figure}
\noindent
The intragap soliton levels are analyzed in more detail
in Fig. \ref{F:level}.
When the coupling strength increases without any Coulomb interaction, the
mid-gap level due to a neutral soliton keeps still, while the
charged-soliton levels deviate from the gap center.
symmetry.
Once the Coulomb interaction is switched on, the neutral-soliton level
also begins to move away from the gap center breaking the spin up-down
Soliton-related electron and hole levels are necessarily symmetric with
respect to the center of the gap in $M\!X$ chains, while such a symmetry
is generally broken in $M\!M\!X$ chains.
The electron-hole symmetry is kept and broken in the $M\!X$ Hamiltonian
(\ref{E:HMX}) and the $M\!M\!X$ Hamiltonian (\ref{E:HMMX}), respectively.

   Now we show in Figure \ref{F:OC} photoinduced absorption spectra which
are expected of $M\!X$ and $M\!M\!X$ chains.
Since photoinduced solitons are necessarily in pairs, we superpose
the S$^-$ (S$^{0\downarrow}$) spectrum on the S$^+$ (S$^{0\uparrow}$)
spectrum.
The spectra of S$^+$ and S$^-$ are degenerate for $M\!X$ chains but
distinguishable for $M\!M\!X$ chains.
{\it The asymmetry between the S$^\pm$ absorption bands can remain even in
the absence of the Coulomb correlation provided the electron-phonon
interaction is sufficiently strong.}
Although the spectra of S$^{0\uparrow}$ and S$^{0\downarrow}$ are
degenerate for both $M\!X$ and $M\!M\!X$ chains, those for $M\!M\!X$
chains are interesting in themselves.
The S$^0$ spectrum is single-peaked for $M\!X$ chains but splits into two
definite peaks for $M\!M\!X$ chains.
{\it The doublet structure of the S$^0$ spectrum is due to the combination
of the broken electron-hole symmetry and relevant Coulomb correlation.}
In principle $M\!X$ chains also lose the electron-hole symmetry with their
$X$ $p_z$ electrons activated \cite{G10566}, which indeed applies to the
Ni$X$ compounds \cite{O8438}.
However, it is not the case with the Pt$X$ and Pd$X$ compounds, where
the energy level of the $X$ $p_z$ orbitals is much lower than that of the
$M$ $d_{z^2}$ orbitals and thus the $p$-orbital contribution may
effectively be incorporated into the intermetal supertransfer energy of
the single-band Hamiltonian (\ref{E:HMX}).
In fact a two-band calculation of the PtCl compound \cite{W6435} still
results in a single-peaked S$^0$ absorption band near the gap center.

\section{Summary and Discussion}

   Since polarons are generated only from electron-hole pairs with large
excess energies, lower-energy-excited charge transfer excitons relax into
soliton pairs or decay by luminescence \cite{O2023}.
A neutral-soliton pair has a lower formation energy
\cite{G6408,Y189,T2212,I1088} but a shorter life time \cite{O2023} under
realistic parametrizations.
Photoexcited
[Pt(en)$_2X$](ClO$_4$)$_2$ ($X=\mbox{Cl},\mbox{Br}$) \cite{K2237,O861} and
[Pt(en)$_2$I](ClO$_4$)$_2$ \cite{O6330} indeed yield intragap absorptions
attributable to neutral and charged solitons, respectively.
The charged-soliton spectrum has a single-peaked Gaussian shape, which is
well consistent with Fig. \ref{F:OC}, while the neutral-soliton spectrum
consists of a midgap main band and an accompanying shoulder structure,
which demands further driving forces such as lattice fluctuations
\cite{I1380}.

   Solitons may be photogenerated in $M\!M\!X$ complexes as well, yielding
a contrastive absorption spectrum of doublet structure.
The separation between the two S$^0$ bands is in proportion to the on-site
Coulomb repulsion and may therefore be reduced with pressure applied,
whereas the asymmetry between the two S$^\pm$ bands is sensitive to the
electron-lattice coupling and may thus be enhanced by the halogen
replacement $\mbox{Cl}\rightarrow\mbox{Br}\rightarrow\mbox{I}$.
Photoinduced absorption measurements on $M\!M\!X$ complexes may not only
reveal the novel doublet structures of soliton-induced spectra but also
give a key to the unsettled problem in $M\!X$ complexes$-$the shoulder
structure of the S$^0$ absorption spectrum.
Let us make a close collaboration between experimental and theoretical
investigations of dynamic properties of $M\!M\!X$ complexes.

\acknowledgments

   The authors are grateful to K. Iwano and H. Okamoto for fruitful
discussions and helpful comments.
This work was supported by the Ministry of Education, Culture, Sports,
Science, and Technology of Japan, the Nissan Science Foundation, and
the Iketani Science and Technology Foundation.

\widetext

\begin{references}

\bibitem{G6408}
   J. T. Gammel, A. Saxena, I. Batisti\'c, A. R. Bishop, and
   S. R. Phillpot,
      Phys. Rev. B {\bf 45}, 6408 (1992).

\bibitem{O2023}
   H. Okamoto and M. Yamashita,
      Bull. Chem. Soc. Jpn. {\bf 71}, 2023 (1998).

\bibitem{N3865}
   K. Nasu,
      J. Phys. Soc. Jpn. {\bf 52}, 3865 (1983).

\bibitem{B13228}
   I. Batisti\'c, J. T. Gammel, and A. R. Bishop,
      Phys. Rev. B {\bf 44}, 13228 (1991);
   H. R\"oder, A. R. Bishop, and J. T. Gammel,
      Phys. Rev. Lett. {\bf 70}, 3498 (1993).

\bibitem{O9}
   H. Okamoto, T. Mitani, K. Toriumi, and M. Yamashita,
      Mater. Sci. Eng. B {\bf 13}, L9 (1992).

\bibitem{G1191}
   J. T. Gammel and G. S. Kanner,
      Synth. Met. {\bf 70}, 1191 (1995).

\bibitem{S1698}
   W. P. Su, J. R. Schrieffer, and A. J. Heeger,
      Phys. Rev. Lett. {\bf 42}, 1698 (1979);
      Phys. Rev. B {\bf 22}, 2099 (1980).

\bibitem{T2388}
   H. Takayama, Y. R. Lin-Liu, and K. Maki,
      Phys. Rev. B {\bf 21}, 2388 (1980).

\bibitem{I137}
   S. Ichinose:
      Solid State Commun. {\bf 50} (1984) 137.

\bibitem{O250}
   Y. Onodera:
      J. Phys. Soc. Jpn. {\bf 56} (1987) 250.

\bibitem{B339}
   D. Baeriswyl and A. R. Bishop,
      J. Phys. C {\bf 21}, 339 (1988).

\bibitem{K2122}
   N. Kuroda, M. Sakai, Y. Nishina, M. Tanaka, and S. Kurita,
      Phys. Rev. Lett. {\bf 58}, 2122 (1987).

\bibitem{O2248}
   H. Okamoto, T. Mitani, K. Toriumi and, M. Yamashita,
      Phys. Rev. Lett. {\bf 69}, 2248 (1992).

\bibitem{K435}
   M. Kuwabara and K. Yonemitsu,
      Mol. Cryst. Liq. Cryst. {\bf 341}, 533 (2000);
                              {\bf 343}, 47 (2000);
      J. Phys. Chem. Solids {\bf 62}, 435 (2001).

\bibitem{Y125124}
   S. Yamamoto,
      Phys. Lett. A {\bf 258}, 183 (1999);
      J. Phys. Soc. Jpn. {\bf 69}, 13 (2000);
      Phys. Rev. B {\bf 63}, 125124 (2001).

\bibitem{K10068}
   H. Kitagawa, N. Onodera, T. Sonoyama, M. Yamamoto, T. Fukawa,
   T. Mitani, M. Seto, and Y. Maeda,
      J. Am. Chem. Soc. {\bf 121}, 10068 (1999).

\bibitem{Y1198}
   S. Yamamoto,
      J. Phys. Soc. Jpn. {\bf 70}, 1198 (2001).

\bibitem{S1405}
   B. I. Swanson, M. A. Stroud, S. D. Conradson and M. H. Zietlow,
      Solid State Commun. {\bf 65}, 1405 (1988);
   M. A. Stroud, H. G. Drickamer, M. H. Zietlow, H. B. Gray, and
   B. I. Swanson,
      J. Am. Chem. Soc. {\bf 111}, 66 (1989).

\bibitem{Y140102}
   S. Yamamoto,
      Phys. Rev. B {\bf 64}, 140102(R) (2001).

\bibitem{M046401}
   H. Matsuzaki, T. Matsuoka, H. Kishida, K. Takizawa, H. Miyasaka,
   K. Sugiura, M. Yamashita, and H. Okamoto,
      Phys. Rev. Lett. {\bf 90}, 046401 (2003).

\bibitem{K1931}
   H. Kitagawa, N. Onodera, J.-S. Ahn, and T. Mitani,
      Synth. Met. {\bf 86}, 1931 (1997).

\bibitem{Y189}
   S. Yamamoto and M. Ichioka,
      J. Phys. Soc. Jpn. {\bf 71}, 189 (2002);
   S. Yamamoto,
      Mol. Cryst. Liq. Cryst. {\bf 379}, 555 (2002).

\bibitem{J1415}
   S. Jin, T. Ito, K. Toriumi, and M. Yamashita,
      Acta Cryst. C {\bf 45}, 1415 (1989).

\bibitem{K40}
   N. Kimura, H. Ohki, R. Ikeda, M. Yamashita,
      Chem. Phys. Lett. {\bf 220}, 40 (1994).

\bibitem{W1195}
   Y. Wada, T. Furuta, M. Yamashita, and K Toriumi,
      Synth. Met. {\bf 70}, 1195 (1995).

\bibitem{K2163}
   M. Kuwabara and K. Yonemitsu,
      J. Mater. Chem. {\bf 11}, 2163 (2001).

\bibitem{B1155}
   L. G. Butler, M. H. Zietlow, C.-M. Che, W. P. Schaefer,
   S. Sridhar, P. J. Grunthaner, B. I. Swanson, R. J. H. Clark, and
   H. B. Gray,
      J. Am. Chem. Soc. {\bf 110}, 1155 (1988).

\bibitem{B4562}
   S. A. Borshch, K. Prassides, V. Robert, and A. O. Solonenko,
      J. Chem. Phys. {\bf 109}, 4562 (1998).

\bibitem{B444}
   C. Bellitto, A. Flamini, L. Gastaldi, and L. Scaramuzza,
      Inorg. Chem. {\bf 22}, 444 (1983).

\bibitem{B2815}
   C. Bellitto, G. Dessy, and V. Fares,
      Inorg. Chem. {\bf 24}, 2815 (1985).

\bibitem{Y422}
   S. Yamamoto,
      Phys. Lett. A {\bf 247}, 422 (1998);
      Synth. Met. {\bf 103}, 2683 (1999).

\bibitem{I1380}
   K. Iwano and K. Nasu,
      J. Phys. Soc. Jpn. {\bf 61}, 1380 (1992).

\bibitem{T2212}
   Y. Tagawa and N. Suzuki,
      J. Phys. Soc. Jpn. {\bf 64}, 2212 (1995).

\bibitem{C4604}
   C.-M. Che, F. H. Herbstein, W. P. Schaefer, R. E. Marsh, and
   H. B. Gray,
      J. Am. Chem. Soc. {\bf 105}, 4604 (1983).

\bibitem{C409}
   R. J. H. Clark, M. Kurmoo, H. M. Dawes, and M. B. Hursthouse,
      Inorg. Chem. {\bf 25}, 409 (1986).

\bibitem{Y165113}
   S. Yamamoto,
      Phys. Rev. B {\bf 66}, 165113 (2002).

\bibitem{T4074}
   Y. Tagawa and N. Suzuki,
      J. Phys. Soc. Jpn. {\bf 59}, 4074 (1990).

\bibitem{G10566}
   J. T. Gammel, R. J. Donohoe, A. R. Bishop, and B. I. Swanson,
      Phys. Rev. B {\bf 42}, 10566 (1990).

\bibitem{O8438}
   H. Okamoto, Y. Shimada, Y. Oka, A. Chainani, T. Takahashi, H. Kitagawa,
   T. Mitani, K. Toriumi, K. Inoue, T. Manabe, and M. Yamashita,
      Phys. Rev. B {\bf 54}, 8438 (1996).

\bibitem{W6435}
   S. W. Weber-Milbrodt, J. T. Gammel, A. R. Bishop, and
   E. Y. Loh, Jr.,
      Phys. Rev. B {\bf 45}, 6435 (1992).

\bibitem{I1088}
   K. Iwano,
      J. Phys. Soc. Jpn. {\bf 66}, 1088 (1997).

\bibitem{K2237}
   N. Kuroda, M. Ito, Y. Nishina, and M. Yamashita,
      J. Phys. Soc. Jpn. {\bf 62}, 2237 (1993).

\bibitem{O861}
   H. Okamoto, Y. Kaga, Y. Shimada, Y. Oka, Y. Iwasa, T. Mitani, and
   M. Yamashita,
      Phys. Rev. Lett. {\bf 80}, 861 (1998).

\bibitem{O6330}
   H. Okamoto, Y. Oka, T. Mitani, and M. Yamashita,
      Phys. Rev. B {\bf 55}, 6330 (1997).

\end{references}
\end{document}